\documentclass[a4paper,11pt]{article}
\pdfoutput=1 

\usepackage{jcappub} 
\usepackage[T1]{fontenc} 
\usepackage{hyperref,caption}
\usepackage[flushleft]{threeparttable}

\newcommand{\ceffs}{c_\mathrm{eff}^2}
\newcommand{\cviss}{c_\mathrm{vis}^2}

\title{\boldmath Updating non-standard neutrinos properties with Planck-CMB data and full-shape analysis of BOSS and eBOSS galaxies}

\author[a]{Suresh Kumar,}
\author[b,c]{Rafael C. Nunes,}
\author[a]{Priya Yadav}
\affiliation[a]{Department of Mathematics, Indira Gandhi University, Meerpur, Haryana 122502, India}
\affiliation[b]{Instituto de F\'{i}sica, Universidade Federal do Rio Grande do Sul, 91501-970 Porto Alegre RS, Brazil}
\affiliation[c]{Divis\~{a}o de Astrof\'{i}sica, Instituto Nacional de Pesquisas Espaciais, Avenida dos Astronautas 1758, S\~{a}o Jos\'{e} dos Campos, 12227-010, S\~{a}o Paulo, Brazil}

\emailAdd{suresh.math@igu.ac.in}
\emailAdd{rafadcnunes@gmail.com}
\emailAdd{priya.math.rs@igu.ac.in} 

\abstract{Using the latest observational data from Planck-CMB and its combination with the pre-reconstructed full-shape (FS) galaxy power spectrum measurements from the BOSS DR12 sample and eBOSS LRG DR16 sample, we report the observational constraints on the cosmic neutrino properties given by the extended $\Lambda$CDM scenario: $\Lambda$CDM + $N_{\rm eff}$ + $\sum m_{\nu}$ + $c^2_{\rm  eff}$ + $c^2_{\rm  vis}$ + $\xi_{\nu}$, and its  particular case $\Lambda$CDM + $c^2_{\rm  eff}$ + $c^2_{\rm  vis}$ + $\xi_{\nu}$,  where $N_{\rm eff}$, $\sum m_{\nu}$, $c^2_{\rm  eff}$, $c^2_{\rm  vis}$, $\xi_{\nu}$ are the effective number of species, the total neutrino mass, the sound speed in the neutrinos rest frame, the viscosity parameter and the degeneracy parameter quantifying a cosmological leptonic asymmetry, respectively.  We observe that the combination of FS power spectrum measurements with the CMB data significantly improves the parametric space of the models compared to the CMB data alone case. We find no evidence for neutrinos properties other than the ones predicted by the standard cosmological theory. Our most robust observational constraints are given by  CMB + BOSS analysis. For the generalized extended $\Lambda$CDM scenario, we find $c^2_{\rm eff}=0.3304^{+0.0064}_{-0.0075}$, $c^2_{\rm vis}=0.301^{+0.037}_{-0.033}$, $\xi_{\nu} < 0.05$, $N_{\rm eff}=2.90 \pm 0.15$ at 68\% CL, with $\sum m_{\nu} < 0.116$ eV at 95\% CL. These are the strongest limits ever reported for these extended $\Lambda$CDM scenarios.}

\begin{document}
\maketitle
\flushbottom

\section{Introduction}
\label{sec:intro}

The cosmic neutrinos are known to decouple from the rest of the cosmic plasma at $k_B T \sim$ MeV. These freely streaming relic neutrinos lead to the cosmic neutrino background (CNB) just like the directly observed cosmic microwave background (CMB) of cosmic photons. Though the relic neutrinos background is yet to be detected directly, some indirect measures have been established in this regard by using CMB, viz., estimates from the primordial abundances of light elements, clustering of the large scale structure (LSS), as well as few other cosmological and astrophysical observations (see \cite{Dolgov_2002,LESGOURGUES_2006,Lattanzi} for a review). On the other hand, neutrino oscillations measured at terrestrial experiments indicate that at least two massive neutrinos exist in nature \cite{Athar_2022}. The KATRIN experiment  provides an upper limit of 0.9 eV on the neutrino mass scale \cite{katrin} while the neutrino mass hierarchy has recently been discussed in \cite{S_Gariazzo,R_Jimenez}. The properties of the neutrinos cause direct effects in important cosmological sources, and thereby play an important role in the dynamics of the Universe, and in the determination of cosmological parameters \cite{Dolgov_2002,LESGOURGUES_2006,Lattanzi}. The relic neutrinos may cause only gravitational effects on the CMB and LSS because they are decoupled (free-streaming particles) well before the time of recombination and structure formation.

The standard parameters that characterize the neutrinos effects on cosmological sources are the effective number of species $N_{\rm eff}$ and the total neutrino mass $\sum m_{\nu}$.
The joint analysis with CMB and baryonic acoustic oscillations (BAO) data places an upper bound at 95\% confidence level (CL) on the total neutrino mass and number, viz., $\sum m_{\nu} < 0.12$ eV and $N_{\rm eff}= 3.04 \pm 0.33$, respectively \cite{planck2020}. The most robust and  recent upper bound at 95\% CL on the total neutrino mass is $\sum m_{\nu} < 0.09$ eV \cite{Di_Valentino_2021_neutrinos}, obtained within the minimal $\Lambda$CDM framework. In \cite{Chudaykin_neutrinos}, the authors report a 4$\sigma$ evidence for nonzero neutrino mass. On the other hand, the value of $N_{\rm eff}$ via theoretical calculations is well determined within the framework of the  standard model, viz., $N_{\rm eff}=3.046$. The evidence for a deviation from this value can be a signal that the radiation content of the Universe is not only due to photons and neutrinos, but also due to some extra relativistic relics, the so-called dark radiation sector. A larger value for $N_{\rm eff}$ could also arise from different physics, for instance, axions, decay of non-relativistic,  gravity waves, dark energy, and  a few other phenomena. The effects of $N_{\rm eff}$ are investigated in several contexts beyond the $\Lambda$CDM cosmology \cite{Kumar:2016zpg,Yang:2019uog,Yang:2020tax}, and also with regard to a possible solution for the $H_0$ and/or $S_8$ tensions \cite{snowmass_r,Di_Valentino_2021_S8, Kreisch_2020,Brinckmann_2021,Das,Vagnozzi_2020_H0,Anchordoqui_2021,Alcaniz,SRC_2021}.

On the other hand, the information on the dark relativistic background can be obtained not only from its effects on the expansion rate of the Universe but also from its clustering properties. Going beyond the standard properties, two phenomenological parameters $c^2_{\rm eff}$ and $c^2_{\rm vis}$ can be introduced. Here, $c^2_{\rm eff}$ is the sound speed in the neutrinos rest frame and $c^2_{\rm vis}$ is the viscosity parameter, which parameterizes the anisotropic stress. The evolution of standard neutrinos (non-interacting free-streaming neutrinos) is obtained for $c^2_{\rm eff}=c^2_{\rm vis}=1/3$.  It is important to mention that these parametrizations were strongly inspired by pioneer works about dark matter properties \cite{Hu_1998,Hu_1998_b}. These perturbation parameters can be constrained through measurements of the CMB anisotropies and LSS observations since dark radiation is coupled through gravity with all the remaining components. Observational constraints on $c^2_{\rm eff}$ and $c^2_{\rm vis}$ are investigated via different methods and approaches \citep{Archidiacono_2012,Archidiacono_2013,Gerbino_2013, Sellentin_2015,Audren_2015,Nunes_2017,Oldengott_2015,Forastieri_2015}. Measuring a deviation from $(c^2_{\rm eff},c^2_{\rm vis})=1/3$ can  open a new window for testing the dark radiation component, since, for example, a deviation in $c^2_{\rm vis}$ could indicate a possible non-standard interaction.

Another natural extension of the physics properties of the neutrino is to consider  a certain degree of lepton asymmetry (a cosmological leptonic asymmetry), which is usually parametrized by the so-called degeneracy parameter $\xi_{\nu}=u_{\nu}/T_{\nu0}$ \citep{Lesgourgues_1999,Serpico_2005,Simha_2008}, where $u_{\nu}$ is the neutrino chemical potential and $T_{\nu0}$ is the current temperature of the relic neutrinos background, $T_{\nu0}/T_{CMB} = (4/11)^{1/3}$. The leptonic asymmetry also shifts the equilibrium between protons and neutrons at the Big Bang Nucleosynthesis (BBN) epoch, leading to indirect effects on the CMB anisotropy through the primordial helium abundance $Y_{He}$. The effects of a leptonic asymmetry on BBN and CMB are investigated in many studies \cite{Caramete_2014,Abazajian_2002,Schwarz_2013,Lattanzi_2005,Kinney_1999,Mangano_2012,Castorina_2012,Oldengott_2017,Barenboim_2017,Bonilla_2019,Gelmini_2020,Zeng_2019,Grohs_2017,Kawasaki2022,Burns_2022}. Also, it has been argued that the $H_0$ tension can be slightly alleviated up to a significant level via a non-null cosmological leptonic asymmetry \cite{Barenboim_2017,Seto_2021,Yeung_2021}.

In addition to the effects on CMB, non-standard neutrinos properties can also affect the LSS of the Universe. In recent years, beyond the information contained within the (reconstructed) BAO peaks, significant efforts have been made for extracting LSS clustering information from the full-shape (FS) power spectrum of biased tracers of the LSS. Significant advances in the so-called Effective Field Theory of LSS (EFTofLSS) have led to such efforts and concrete applications of real data from the galaxy redshift surveys ~\cite{Baumann:2010tm,Carrasco:2012cv,Pajer:2013jj,Carroll:2013oxa,Senatore:2014via,Senatore:2014vja,Baldauf:2015aha,Foreman:2015lca,Senatore:2017hyk,Nishimichi:2020tvu,Smith:2022,Simon:2022}.
The effective field theory-based full-shape analysis of the power spectrum has been applied to derive constraints on $\sum m_{\nu}$ and $N_{\rm eff}$ in \cite{Chudaykin_neutrinos,Ivanov:2019hqk,Philcox:2020vvt,Ivanov:2021zmi}. Also, EFTofLSS has been used to constrain the standard cosmological parameters and models beyond the $\Lambda$CDM cosmology \cite{Philcox:2021kcw,Philcox_22,Colas:2019ret,DAmico:2019fhj,Ivanov:2019pdj,Philcox:2020vvt,Cabass_22,DAmico:2020kxu,Ivanov:2020ril,DAmico:2020ods,Philcox:2020xbv,Chudaykin:2020hbf,Chudaykin:2020ghx,DAmico:2020tty,Nunes_fs_2021_a,Simon2021a,Lague:2021frh,Sgier,Ivanov:2021fbu,Oliver_Philcox_2021,Sunny_2022,pedro_2022,Zhang:2022,AP:2022,Rizzo:2022}. Besides the EFTofLSS, a number of other theoretical modeling approaches have been adopted in the analysis for galaxy clustering data to derive constraints on $\sum m_{\nu}$ and $N_{\rm eff}$ (see \cite{Chiang_2019,Bayer_2021,Boyle_2021,Chen_2021,Brinckmann_2019_nu,Banerjee_2020,Cuesta:2015iho,Castorina_2015,Giusarma_2016_nu1,Vagnozzi_2021_nu1,Vagnozzi:2017ovm,Ivanov:2021fbu,Giusarma:2018jei,Zennaro_2018,Giar2021,Xu_2021,Loureiro_2019,SV_2022} for a short list).

In this work, our goal  is to make use of the EFTofLSS approach to explore whether redshift-space galaxy clustering data can improve state-of-the-art constraints on the non-standard neutrinos properties $c^2_{\rm eff}$, $c^2_{\rm vis}$ and $\xi_{\nu}$. In this paper, we consider two extensions of the minimal $\Lambda$CDM model: i) $\Lambda$CDM + $c^2_{\rm  eff}$ + $c^2_{\rm  vis}$ + $\xi_{\nu}$ and ii) $\Lambda$CDM + $N_{\rm eff}$ + $\sum m_{\nu}$ + $c^2_{\rm  eff}$ + $c^2_{\rm  vis}$ + $\xi_{\nu}$.  The paper is organized as follows. In Section \ref{model}, we briefly summarize the theoretical framework adopted in this work. In Section \ref{data}, we describe the datasets and analysis methodology. In Sections \ref{results} and \ref{conclusions}, we present our main results and conclusions, respectively.

\section{Theoretical Model}
\label{model}

Neutrinos and antineutrinos of each flavour $\nu_i$ ($i = e,\mu,\tau$) behave like relativistic particles in the very early Universe. The energy density and pressure of massive degenerate neutrinos and antineutrinos (one species) are given by

\begin{eqnarray}
 \rho_{\nu_i} +  \rho_{\bar{\nu_i}} = T^4_{\nu} \int \frac{d^3q}{2(\pi)^3} q^2 E_{\nu_i} (f_{\nu_i}(q) + f_{\bar{\nu_i}})) 
\end{eqnarray}
and 

\begin{eqnarray}
3 (p_{\nu_i} +  p_{\bar{\nu_i}}) = T^4_{\nu} \int \frac{d^3q}{2(\pi)^3} \frac{q^2}{E_{\nu_i}}(f_{\nu_i}(q) + f_{\bar{\nu_i}})).
\end{eqnarray}
Here we have used $\hbar = c = k_B=1$. Further, $E^2_{\nu_i} = q^2 + a^2 m_{\nu_i}$ is one flavour neutrino/antineutrinos energy and $q = a p$ is the comoving momentum. The functions $f_{\nu_i}(q)$, $f_{\bar{\nu_i}}$ are the Fermi-Dirac phase space distributions given by

\begin{equation}
f_{\nu_i}(q) = \frac{1}{e^{E_{\nu_i}/T_{\nu} - \xi_{\nu}} + 1},\;\; f_{\bar{\nu_i}}(q) = \frac{1}{e^{E_{\bar{\nu_i}}/T_{\nu} - \xi_{\bar{\nu}}} + 1},  
\end{equation}
where $\xi_{\nu} = \mu/T_{\nu}$ is the neutrino degeneracy parameter. Usually, in cosmological analysis, $\xi_{\nu}$ is fixed  to 0, but the presence of a significant and non-null $\xi_{\nu}$ can have some  cosmological implications \cite{Caramete_2014,Abazajian_2002,Schwarz_2013,Lattanzi_2005,Kinney_1999,Mangano_2012,Castorina_2012,Oldengott_2017,Barenboim_2017,Bonilla_2019,Kawasaki2022,Gelmini_2020,Zeng_2019,Grohs_2017,Seto_2021,Yeung_2021}. A large $\xi_{\nu}$ implies large neutrino asymmetry, which may contradict popular leptogenesis scenarios in which sphalerons effectively transfer lepton asymmetry to baryons in the early Universe \cite{1986PhLB..174...45F}. Observational information on $\xi_{\nu}$ therefore has important implications on theories of matter-anti-matter asymmetry in the Universe \cite{Mangano_2011}. If neutrino is Majorana, then $\xi_{\nu} = 0$ \cite{Mangano_2012}. Therefore, if $\xi_{\nu} \neq 0$ is observed, neutrinos must be Dirac particles, and then the measurement of $\xi_{\nu}$ is of fundamental importance. On the other hand, a finite neutrino chemical potential would affect the neutrino energy density, which in turn would modify the CMB anisotropy spectra. The major impact of $\xi_{\nu}$ on CMB physics is that it modifies the expansion rate at early times. 

The effect of $\xi_{\nu}$ can be expressed as an excess in $N_{\rm eff}$, viz.,

\begin{equation}
\label{delta_neff_xi}
\Delta N_{\rm eff} = \sum_i \frac{30}{7} \Big(\frac{\xi_{\nu,i}}{\pi} \Big)^2 + \frac{15}{7}  \Big( \frac{\xi_{\nu,i}}{\pi} \Big)^4,
\end{equation}
where the label $i$ runs over the mass eigenstates. The above equation is only valid for massless neutrinos. With non-zero masses, the thermal distribution and energy density of neutrinos/antineutrinos are modified (see, for example, Ref. \cite{Lesgourgues_1999}). We have not taken into account such modifications in this work. However, as long as the masses of light mass eigenstates are much smaller than their momentum around the epoch of CMB decoupling, the above equation provides a good enough approximation. From Eq. (\ref{delta_neff_xi}), for $\xi_{\nu}> 0$, this fact induces an increase in the expansion rate at early times (in comparison with the hypothesis $\xi_{\nu}=0$ ), and consequently an effect on $H_0$. We will discuss the consequence of this effect later.
\\

In the literature, several approximations for massive neutrino are presented \cite{Lesgourgues_2011_class_neutrinos, Ma_1995,Shoji_2010,Wong_2015,Bird_2013,Blas_2014,Thomas_2018,TT_2018}. Here, we describe the evolution of linear perturbations using the methodology of \cite{Audren_2015}, where an extension for massless and massive neutrinos is described where the null hypothesis with respect to the
parameters $c^2_{\rm eff}$ and $c^2_{\rm vis}$ is assumed to be different from $c^2_{\rm eff}=c^2_{\rm vis}=1/3$. For the massless case, i.e., in the relativistic limit, the continuity, Euler, and shear equations in the synchronous gauge are respectively given by

\begin{eqnarray}
\dot{\delta}_\nu &=&
\left( 1 - 3 \ceffs \right) \frac{\dot{a}}{a}
\left(\delta_\nu+\frac{4}{k^2} \frac{\dot{a}}{a} \theta_\nu \right)
-\frac{4}{3} (\theta_\nu + \frac{\dot{h}}{2})\,,
\\
\dot{\theta}_\nu &=&
\frac{k^2}{4} (3\ceffs) \left(\delta_\nu+\frac{4}{k^2} \frac{\dot{a}}{a} \theta_\nu\right)
- \frac{\dot{a}}{a} \theta_\nu - k^2 \sigma_\nu\,,
\end{eqnarray}

\begin{equation}
\dot{F}_{\nu2} = 2 \dot{\sigma}_\nu = (3  \cviss) \frac{8}{15} (\theta_\nu +  (\dot{h}+6\dot{\eta})/2) - \frac{3}{5} k F_{\nu 3}\,, \label{sigdot_relativistic}
\end{equation}
where $F_{\nu \ell}$ are the Legendre multipoles of the momentum integrated neutrino distribution function as defined in  \cite{Ma_1995}. Further equations in the hierarchy are left unchanged.

For massive neutrinos, the modified Boltzmann hierarchy reads

\begin{eqnarray}
\dot{\Psi}_0
&=& \frac{\dot{a}}{a} \left( 1- 3 c_\mathrm{eff}^2 \right) \frac{q^2}{\epsilon^2} \left[  \Psi_0
+ 3 \frac{\dot{a}}{a}  \frac{5p-\tilde{p}}{\rho+p} \frac{\epsilon }{k q} \Psi_1 \right]
- \frac{qk}{\epsilon} \Psi_1 + \frac{\dot{h}}{6} \frac{d \ln f_0}{d \ln q}~,\\
\dot{\Psi}_1 &=& c_\mathrm{eff}^2 \frac{q k}{\epsilon} \left[ \Psi_0 + 3 \frac{\dot{a}}{a} \frac{5p-\tilde{p}}{\rho+p} \frac{\epsilon}{qk} \Psi_1 \right] - \frac{\dot{a}}{a} \frac{5p-\tilde{p}}{\rho+p} \Psi_1 - \frac{2}{3} \frac{qk}{\epsilon} \Psi_2~.
\end{eqnarray}

The source term of the shear ($l=2$) is given by

\begin{equation}
\dot{\Psi}_2 = \frac{q k}{5 \epsilon} \left(6 c_\mathrm{vis}^2 \Psi_1 - 3 \Psi_3 \right) - 3 c_\mathrm{vis}^2 \frac{2}{15} (\dot{h}+6\dot{\eta})/2  \frac{d \ln f_0}{d \ln q} \,. \\
\end{equation}

Higher momenta in the Boltzmann hierarchy are left unchanged. In the above equantions, $\tilde{p}$ is the so-called the pseudopressure, $f_0(q)$ is the background phase-space distribution of the momentum $q$ of the particle, $\epsilon$ is the comoving energy of the particle. All these quantities are well defined and discussed in \cite{Lesgourgues_2011_class_neutrinos}.

Note that the standard equation evolution (non-interacting free-streaming neutrinos) is obtained for $c^2_{\rm eff}=c^2_{\rm vis}=1/3$. These latter conditions, in combination with $\xi_{\nu} = 0$, fix the evolution of standard neutrinos in cosmological analysis. The main objective of this work is to relax these three conditions. It is important to mention that the above model is a phenomenological model that parametrizes the behaviour of the neutrinos in a fluid-like fashion in terms of an effective sound speed and an effective viscosity parameter. In principle, strictly speaking, there is no reason to expect that a fundamental theory could be effectively described with a value of $c^2_{\rm eff}$ and $c^2_{\rm vis}$ different from the standard value of 1/3. A more fundamental description of neutrino self-interactions has been investigated in several papers, for instance, see \cite{Forastieri_2015,Pan_2017,Kreisch_2020,Park_2019,Brinckmann_2021,Tram_2017,SRC_2021,JM_2022,SG_2020}. Our aim in this work  could be seen as a first attempt to confront non-standard neutrino properties with the full shape of LSS data, before a more fundamental analysis. Thus, in this work we will adopt the phenomenological model as introduced above.
\\

On the other hand, we are interested in analysing the galaxy power spectrum data. To model the power spectrum we consider the one-loop (tree-level) order with all necessary components, including perturbative corrections, galaxy bias, ultraviolet counterterms, infrared re-summation and stochasticity. The details of the model considered in our analysis are described in \citep{Chudaykin:2020aoj} and references therein, including the conversion from real to redshift-space. Here we only review the main qualitative aspects. The model of the power spectrum multipoles takes the following form (before the effects of infrared re-summation and coordinate transformations):

\begin{eqnarray}
\label{eq: power-spectrum-model}
	P_{g,\ell}(k) = P_{g,\ell}^{\rm tree}(k) + P_{g,\ell}^{\rm 1-loop}(k) + P_{g,\ell}^{\rm ct}(k) + P_{g,\ell}^{\rm stoch}(k),
\end{eqnarray}
where the four terms are: i) The usual linear theory Kaiser multipoles,
scaling as the linear-theory power spectrum $P_{\rm lin}(k)$ at tree-level (linear) galaxy power spectrum. ii) The corresponding one-loop corrections perturbation, scaling as $k^2P_{\rm lin}(k)$ on large scales. iii) The counterterms encapsulate complex short-scale physics that can not be modeled perturbatively in addition to contributions from the Finger-of-God effect, and other degenerate effects, such as higher-derivative biases. These counterterms comprise of a fixed scale dependence which is predicted by EFT. In practice, this leads to three counterterms with free amplitudes $\{c_0, c_2, \tilde{c}\}$, with the third parametrizing next-to-leading order effects from Finger-of-God which can affect larger scales than other non-linearities. iv) The term $P^\mathrm{noise}$ includes stochastic contributions to the galaxy power spectrum which, to one-loop order, includes only a constant, and direction-independent Poissonian shot-noise whose amplitude is a free parameter, $P_\mathrm{shot}$.

In this work, we limit ourselves to the monopole and quadrupole moments ($\ell=$ 0, 2).
All multipoles are computed from the 2D anisotropic galaxy power spectrum as

\begin{eqnarray}
P_{g,\ell}(k) = \frac{2\ell+1}{2} \int_{-1}^{1} d\mu P_g(k,\mu) P_{\ell}(\mu),
\end{eqnarray}
where $\mu$ is cosine of the angle between a Fourier mode $k$ and the line-of-sight
direction $z$, whereas $P_{\ell}(\mu)$ are the Legendre polynomials of order $\ell$. 

Note that  the basis of bias operators, relating the galaxy ($\delta_g$) and matter ($\delta$) overdensity fields, is applied as

\begin{eqnarray}
\label{eq:bias}
\delta_g(\vec x) = b_1\delta(\vec x) + \frac{b_2}{2}\delta^2(\vec x)+b_{\mathcal{G}_2}\mathcal{G}_2(\vec x),
\end{eqnarray}
where $\mathcal{G}_2$ is the tidal field operator. This strictly neglects the additional bias parameter $b_{\Gamma_3}$, which we assume to be null in this work.

In total, we obtain a model with seven nuisance parameters: $\{b_1, b_2, b_{\mathcal{G}_2}, P_\mathrm{shot}, c_0, c_2, \tilde{c}\}$. Since this is based on one-loop perturbation theory, we expect it to be accurate for $k\lesssim0.25$ h/Mpc. Here we only use spectral data in this wavenumber range.

Lastly, we account for the effects of an incorrect fiducial cosmology through the Alcock-Paczynski distortion via the rescalings

\begin{eqnarray*}
\label{eq: coord-rescaling}
	k \to k' &\equiv& k\left[\left(\frac{H_{\rm true}}{H_{\rm fid}}\right)^2\mu^2+\left(\frac{D_{A,\rm fid}}{D_{A,\rm true}}\right)^2(1-\mu^2)\right]^{1/2},\\\nonumber
	\mu \to\mu' &\equiv& \mu\left(\frac{H_{\rm true}}{H_{\rm fid}}\right)\left[\left(\frac{H_{\rm true}}{H_{\rm fid}}\right)^2\mu^2+\left(\frac{D_{A,\rm fid}}{D_{A,\rm true}}\right)^2(1-\mu^2)\right]^{-1/2},
\end{eqnarray*}
where unprimed quantities are measured observationally, and all quantities are evaluated at the sample redshift. Here, $H$ is the Hubble parameter $H(z)$, and $D_A$ is the angular diameter distance $D_A(z)$. The fiducial cosmology, which is used to calibrate the geometric distortion parameters, is assumed with the values $\Omega_{\rm m, fid}$ = 0.31 and $H_0=67.6$ km/s/Mpc.

We emphasize that only the linear power spectrum prediction needs to be modified because it is the only place where the neutrinos physics considered here enters.  Only the expansion, thermal history, and linear evolution of perturbations are affected by the model under consideration. So no changes are required in the standard routines. Our approach may be considered as a conservative one, considering the size of the uncertainties associated with current galaxy clustering measurements. However, it may be noted that this approach has already been used in several other works in the recent literature, where the FS galaxy clustering measurements with the only modifications being the input linear matter power spectra are used to constrain both early and late time new physics scenarios. The other species, namely, baryons, cold dark matter and photons follow the standard evolution, both at the background and perturbation levels. The dark energy is fixed to be a cosmological constant.

\section{Data and Methodology}
\label{data}

In order to derive constraints on the neutrinos parameters model, we use the following datasets.
\begin{itemize}
\item \textbf{CMB}: Measurements of CMB temperature anisotropy and polarization power spectra, as well as their cross-spectra, from the \textit{Planck} 2018 legacy data release. We consider the high-$\ell$ \texttt{Plik} likelihood for TT (in the multipole range $30 \leq \ell \leq 2508$) as well as TE and EE (in the multipole range $30 \leq \ell \leq 1996$), in combination with the low-$\ell$ TT-only ($2 \leq \ell \leq 29$) likelihood based on the \texttt{Commander} component-separation algorithm in pixel space, as well as the low-$\ell$ EE-only ($2 \leq \ell \leq 29$) \texttt{SimAll} likelihood~\cite{Planck:2019nip}. We also include measurements of the CMB lensing power spectrum, as reconstructed from the temperature 4-point function~\cite{Planck:2018lbu}. 

\item \textbf{BOSS}: Measurements of the monopole and quadrupole ($\ell=0$ and $\ell=2$ respectively) of the full-shape power spectrum of the BOSS DR12 galaxies, divided into four independent inputs: two distinct redshift bins at $z_{\rm eff}=0.38$ and $z_{\rm eff}=0.61$, observed in the North and South galactic caps (NGC and SGC), respectively. 

\item \textbf{eBOSS LRG}: Measurements of the monopole and quadrupole ($\ell=0$ and $\ell=2$ respectively) of the full-shape power spectrum of the eBOSS luminous red galaxy sample (DR16 eBOSS LRG) at effective redshift $z_{\rm eff}=0.698$, observed in the NGC and SGC.

\end{itemize}

In our analysis, we use the pre-reconstructed eBOSS LRG power spectrum multipoles, covariance matrices and window functions publicly available in \url{https://svn.sdss.org/public/data/eboss/DR16cosmo/tags/v1_0_1} (see \cite{Gil_Mar_n_2020} for details of the sample). The final DR16 eBOSS LRG catalogue contains both old CMASS BOSS LRG observations in 0.6 < $z$ < 1.0 and the new eBOSS LRG observations within the same redshift range.  This new combined LRG sample is considered independent of the two lowest BOSS DR12 redshift samples. The eBOSS LRG and BOSS DR12 samples slightly overlap in redshift range $0.5 < z < 0.75$. Therefore, when we combine these two samples, we use only the low-$z$ part of the BOSS DR12 sample since the highest redshift part is included in the eBOSS DR16 sample. Although this overlap is quite insignificant and does not present significant correlation between the eBOSS LRG and BOSS LRG samples, as validated in \cite{Alam_2021,Zhao_2021}.

We make use of the FS likelihood, which is publicly available at \url{https://github.com/oliverphilcox/full\_shape\_likelihoods} (see \cite{Philcox:2020vvt}),  in the analyses with both DR16 eBOSS LRG and DR12 BOSS. Theoretical predictions for the relevant observables are obtained using the Boltzmann solver \texttt{CLASS-PT}~\cite{Chudaykin:2020aoj}, which is itself an extension of the Boltzmann solver \texttt{CLASS}~\cite{Blas:2011rf,Lesgourgues:2011re}, and it allows to compute the 1-loop auto- and cross-power spectra for matter fields and biased tracers both in real and redshift spaces, incorporating all the ingredients discussed in Sec.~\ref{model} required for the data comparison. We sample the posterior distributions for the parameters of the model through Monte Carlo Markov Chain (MCMC) methods, using the cosmological sampler \texttt{MontePython}~\cite{Audren:2012wb,Brinckmann:2018cvx}. We assess the convergence of the MCMC chains using the Gelman-Rubin parameter $R-1$~\cite{Gelman:1992zz}, requiring $R-1<0.01$ for the chains to be converged.

The full model for the power spectrum is specified by the following nuisance parameters:

\begin{eqnarray}
\label{eq: nuisance-params}
\{b_1,b_2,b_{\mathcal{G}_2}\} \times \{c_{0}, c_{2}, \tilde c\} \times \{P_{\rm shot} \},
\end{eqnarray}
where the first set encodes galaxy bias (from linear, quadratic, and tidal effects respectively), the second gives the counterterms for the monopole and quadrupole, whilst the final set accounts for the stochastic nature of the density field. Since the BOSS regions have different selection functions and calibrations, we allow the parameters to vary freely in each of the four data chunks. We consider the same for eBOSS data. Note that eBOSS sample has two data chunks. We assume the bias parameter priors as in \cite{Philcox:2020vvt}.

For the analysis with FS data only (next, quantified by BOSS + eBOSS + BBN), we vary the following cosmological parameters:

\begin{eqnarray}
\label{baseline_FS}
\{\omega_b, \omega_{\rm cdm}, h, \ln(10^{10}A_s), n_s \} \times \{c^2_{\rm eff}, c^2_{\rm vis}, \xi_{\nu}  \}.
\end{eqnarray}

For the analyses with the CMB data, the baseline reads:
\begin{eqnarray}
\label{baseline_CMB}
\{\omega_b, \omega_{\rm cdm}, \theta_s, \ln(10^{10}A_s), n_s, \tau\} \times \{N_{\rm eff}, \sum m_{\nu}, c^2_{\rm eff}, c^2_{\rm vis}, \xi_{\nu}  \},
\end{eqnarray}
where the first six cosmological parameters in Eq. (\ref{baseline_CMB}) denote the baryon and cold dark matter physical densities, the angular acoustic scale, the amplitude and tilt of the initial curvature power spectrum at the pivot scale $k = 0.05$/Mpc, and the optical depth to reionization, from which we obtain the derived parameters $H_0$, $\Omega_m$ and $S_8 = \sigma_8 (\Omega_m/0.3)^{0.5}$. The parameters $N_{\rm eff}, \sum m_{\nu}, c^2_{\rm eff}, c^2_{\rm vis}, \xi_{\nu}$, have been described previously. In Eq. (\ref{baseline_FS}), the parameter $h$ is the reduced Hubble constant, $h = H_0/100$. We fix the neutrino mass to the lowest mass allowed by oscillation experiments $\sum m_{\nu} =$ 0.06 eV and when it is let free in specific cases, we impose the flat prior $\sum m_{\nu} \in [0.06, 1]$ eV. All other baseline parameters are free for wide ranges of flat priors. In all analyses, we also impose $\xi_{\nu} > 0$. 
\\

We investigate the observational constraints on the two models: i) $\Lambda$CDM + $c^2_{\rm  eff}$ + $c^2_{\rm  vis}$ + $\xi_{\nu}$, assuming $N_{\rm eff} = 3.046$ and $\sum m_{\nu}=0.06$ eV. ii) $\Lambda$CDM + $N_{\rm eff}$ + $\sum m_{\nu}$ + $c^2_{\rm  eff}$ + $c^2_{\rm  vis}$ + $\xi_{\nu}$. In what follows, we present and discuss the main results.

\section{Main Results}
\label{results}
\begin{figure*}
\begin{center}
\includegraphics[width=15cm]{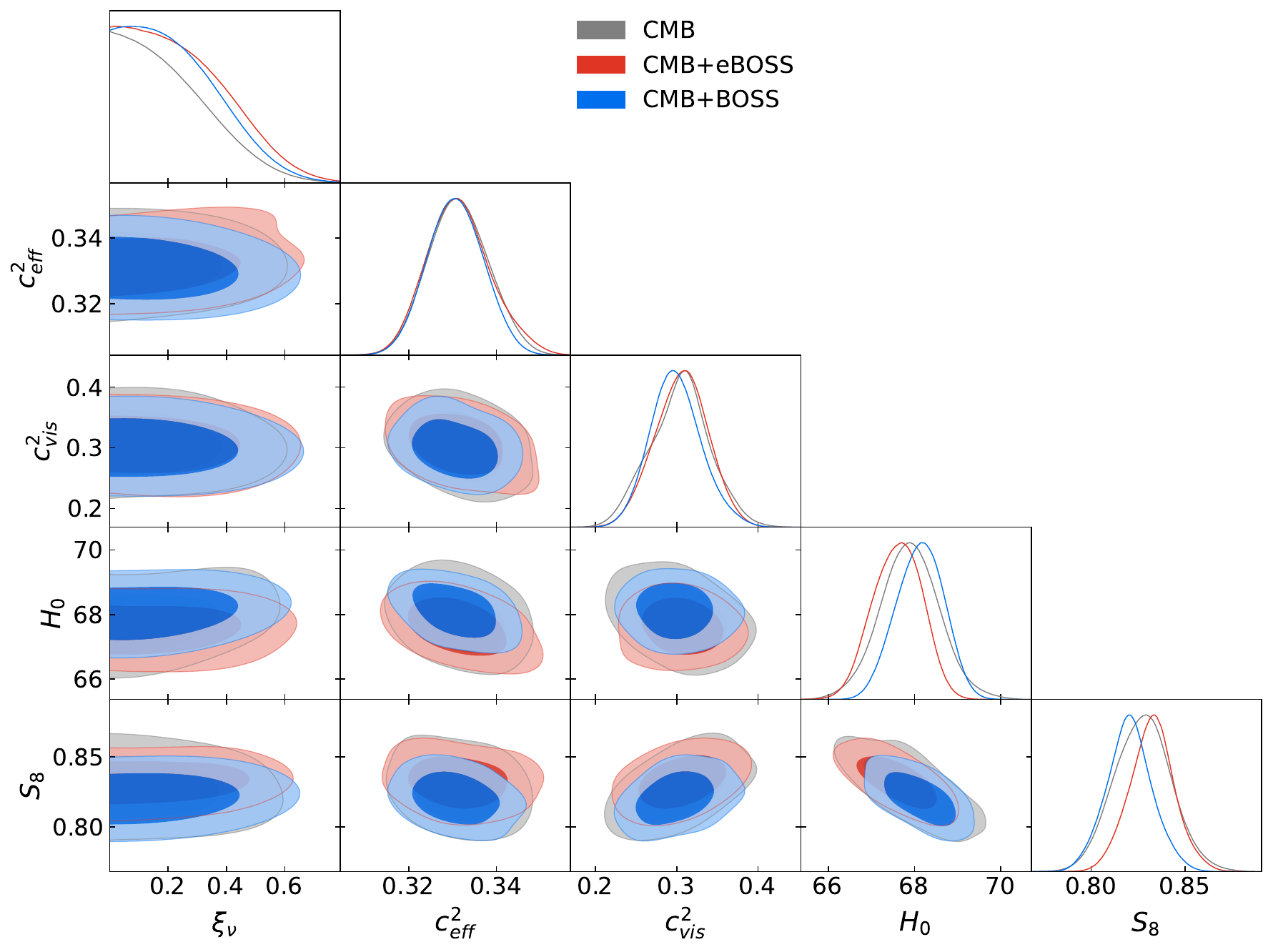} 
\caption{One-dimensional and two-dimensional marginalized confidence regions (68\% and 95\% CL) for $\xi_{\nu}$, $c^2_{\rm  eff}$, $c^2_{\rm  vis}$,  $H_0$ and $S_8$,
obtained from the CMB, CMB+eBOSS, and CMB+BOSS data for the $\Lambda$CDM + $c^2_{\rm  eff}$ + $c^2_{\rm  vis}$ + $\xi_{\nu}$ model.  The parameter $H_0$ is in units of km/s/Mpc.}
\label{fig:cmb_FS}
\end{center}
\end{figure*}

\begin{figure*}
\begin{center}
\includegraphics[width=15cm]{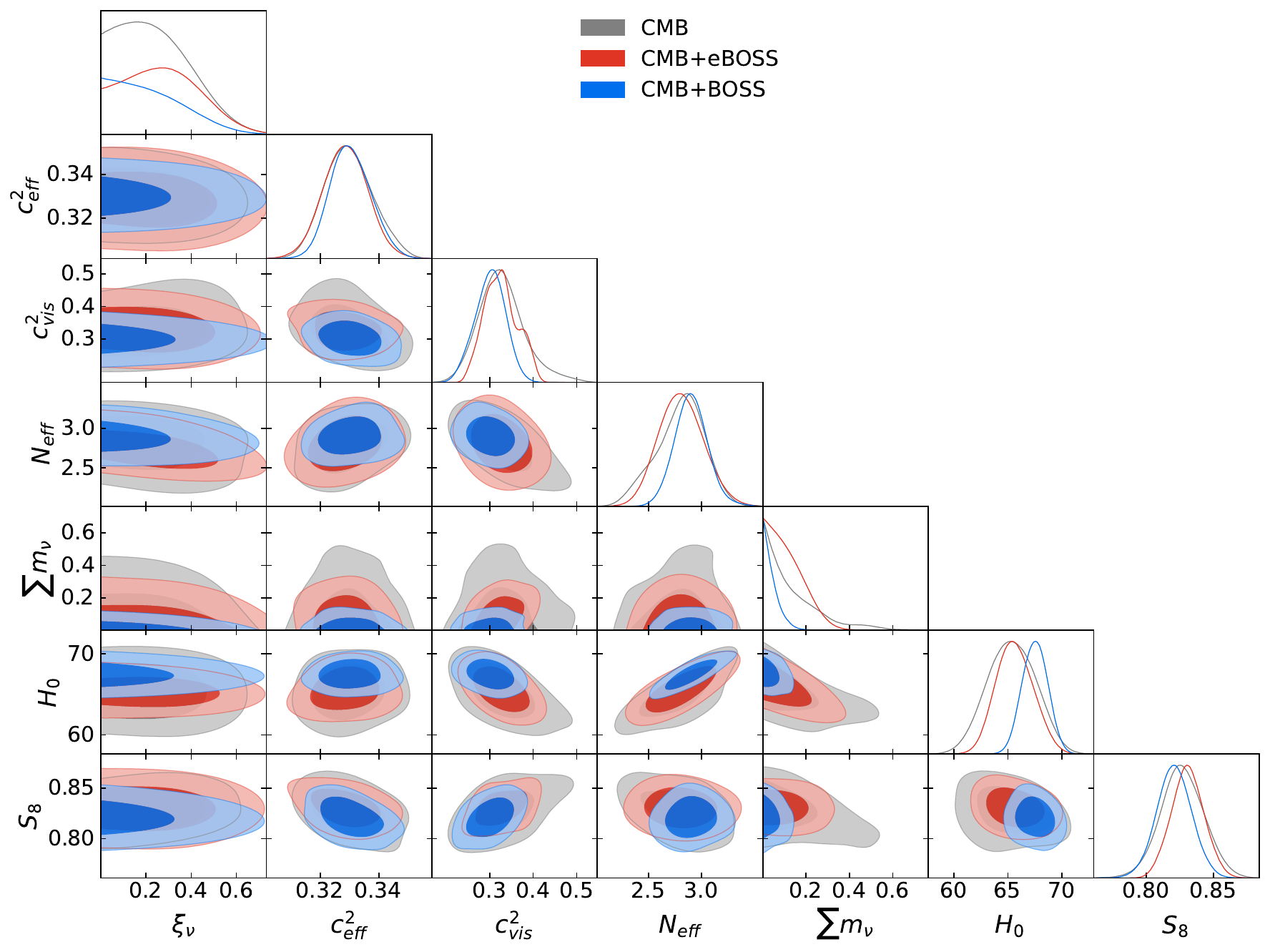} 
\caption{Same as in Figure 1, but for the $\Lambda$CDM + $N_{\rm eff}$ + $\sum m_{\nu}$ + $c^2_{\rm  eff}$ + $c^2_{\rm  vis}$ + $\xi_{\nu}$ model.}
\label{fig:cmb_FS_neff}
\end{center}
\end{figure*}

The constraints on the main parameters of interest of the two models under consideration from CMB, CMB + BOSS and CMB + eBOSS data are summarized in Table \ref{tab:CMB_FS}.  One-dimensional and two-dimensional marginalized confidence regions (68\% and 95\% CL) of the model parameters are displayed in Fig. \ref{fig:cmb_FS} and Fig. \ref{fig:cmb_FS_neff}.

\begin{table*}[hbt!]
\centering
\caption{Constraints at 68\% CL on selected parameters of the two models obtained from CMB, CMB + eBOSS and CMB + BOSS data. The upper limit on neutrino mass scale is reported at 95\% CL.}
\scalebox{0.75}{
\begin{tabular}{|c|c|c|c|c|c|c|cc|cc}       
\hline
Model $\rightarrow$& \multicolumn{3}{|c|}{$\Lambda$CDM + $c^2_{\rm  eff}$ + $c^2_{\rm  vis}$ + $\xi_{\nu}$}  &\multicolumn{3}{|c|}{$\Lambda$CDM + $N_{\rm eff}$ + $\sum m_{\nu}$ + $c^2_{\rm eff}$ +  $c^2_{\rm vis}$ + $\xi_{\nu}$}\\\hline
Parameter & CMB & CMB + eBOSS & CMB + BOSS & CMB & CMB + eBOSS & CMB + BOSS  \\ \hline
$c^2_{\rm eff}$& $0.3313\pm 0.0068          $ & $0.3313^{+0.0065}_{-0.0078}$ &  $0.3306\pm 0.0063          $ &  $0.3297^{+0.0076}_{-0.0090}$ & $0.3285\pm 0.0074          $ & $0.3304^{+0.0064}_{-0.0075}$   \\
$c^2_{\rm vis}$& $0.304\pm 0.037            $ & $0.306\pm 0.033            $ &$0.299^{+0.029}_{-0.033}   $ &$0.328^{+0.043}_{-0.057}   $ &$0.326^{+0.031}_{-0.048}   $&  $0.301^{+0.037}_{-0.033}   $   \\
$\xi_{\nu}$ &  $<0.23  $  & $<0.30            $ &$<0.29     $  & $<0.39                   $  & $< 0.256$ &$< 0.0508                  $    \\
$\sum m_{\nu} \,[{\rm eV}]$& $ 0.06$& $ 0.06$ & $0.06$    &$<0.413    $&$< 0.247$ &$< 0.116                   $ \\
$N_{\rm eff}$ & $ 3.046$ & $ 3.046$ & $3.046$   & $2.80^{+0.26}_{-0.19}      $       &$2.81^{+0.20}_{-0.22}      $ & $2.90\pm 0.15              $   \\
$H_0\,[{\rm km}/{\rm s}/{\rm Mpc}]$&  $67.89\pm 0.68             $ & $67.61\pm 0.55             $ &$68.13\pm 0.54             $ &$65.3\pm 2.2               $& $65.6^{+1.6}_{-1.8}        $ &$67.5\pm 1.2               $  \\
$\Omega_m$                        & $0.3117\pm 0.0079          $ & $0.3161^{+0.0067}_{-0.0080}$ &$0.3086\pm 0.0067          $ &$0.332^{+0.017}_{-0.026}   $& $0.329^{+0.013}_{-0.016}   $ &$0.3098\pm 0.0081          $    \\
$S_8$                          &  $0.828\pm 0.015            $ &  $0.832\pm 0.012            $ &$0.820\pm 0.012            $ &$0.827\pm 0.016            $  & $0.831\pm 0.012            $ &$0.821\pm 0.013            $  \\

\hline                                                
\end{tabular}
}
\label{tab:CMB_FS}
\end{table*}

First we discuss our results obtained from the CMB data only analysis of the two models under consideration. To the authors' knowledge, the parameters $c^2_{\rm eff}$ and $c^2_{\rm vis}$ were constrained in  \cite{Audren_2015} using CMB data only, from the Planck-CMB 2013 data. Thus, our results obtained from the Planck-CMB 2018 legacy data release  provide an update on these non-standard neutrinos properties. The authors in \cite{Audren_2015} report $c^2_{\rm eff} = 0.314 \pm 0.013$, $c^2_{\rm vis}=0.49^{+0.12}_{-0.22}$ at 68\% CL for the $\Lambda$CDM + $c^2_{\rm eff}$ +  $c^2_{\rm vis}$ model and $c^2_{\rm eff} = 0.312 \pm 0.013$, $c^2_{\rm vis}=0.56^{+0.14}_{-0.24}$ at 68\% CL for the  $\Lambda$CDM + $N_{\rm eff}$ + $\sum m_{\nu}$ + $c^2_{\rm eff}$ +  $c^2_{\rm vis}$ model.  Note that we have one additional parameter in our baseline, namely $\xi_{\nu}$. The parameter $\xi_{\nu}$ does not show correlation with $c^2_{\rm eff}$ and $c^2_{\rm vis}$ and this addition does not weaken a direct comparison with previous results, because no significant changes are expected. We can notice a significant improvement on the parameters $c^2_{\rm eff}$ and $c^2_{\rm vis}$ in the analysis with the Planck-CMB 2018 data. We obtain one order of magnitude improvement in the error bars. Also, we do not find any deviations from the standard prediction, and both parameters $c^2_{\rm eff}$ and $c^2_{\rm vis}$ are fully compatible with $c^2_{\rm eff} = c^2_{\rm vis} = 1/3$, while the analysis from Planck-CMB 2013 data shows a deviation at 1$\sigma$ CL in $c^2_{\rm vis}$ \cite{Audren_2015}. Within this general model-baseline, we find $\sum m_{\nu} < 0.413 $ eV at 95\% CL, and no deviation in $N_{\rm eff}$ from its default value.

As well known, assuming the minimal $\Lambda$CDM scenario, Planck-CMB data analysis provides $H_0=67.4 \pm 0.5$ km s$^{-1}$Mpc$^{-1}$ \cite{planck2020}, which is in $\sim$ $5\sigma$ tension with the SH0ES team local measurement $H_0 = 73.30 \pm 1.04$ km s$^{-1}$Mpc$^{-1}$~\cite{Riess_H0}.
Additionally, many other late time measurements are in agreement with a higher value for the Hubble constant (see the discussion in~\cite{Di_Valentino_2021_H0,Perivolaropoulos_H0,snowmass_r}). Motivated by these observational discrepancies, unlikely to disappear completely by introducing multiple and unrelated systematic errors, it has been widely discussed in the literature whether new physics beyond the standard cosmological model can solve the $H_0$ tension (see \cite{Di_Valentino_2021_H0,Perivolaropoulos_H0,snowmass_r} and references therein for a review). A positive value for $\xi_{\nu}$ will induce an increase in the expansion rate at early times, because it increases an excess in $N_{\rm eff}$ (see Eq. (\ref{delta_neff_xi})), and consequently will affect the $H_0$ estimates. Thus, the parameters $\xi_{\nu}$ and $H_0$ are positively correlated, and it has been argued that the $H_0$ tension can be slightly alleviated for a significant value of $\xi_{\nu} > 0$ \cite{Seto_2021,Yeung_2021}. For CMB data only analysis, we note that the two baseline-models analyzed here are not able to alleviate the $H_0$ tension. We find, $H_0=67.89 \pm 0.68$ km/s/Mpc and $H_0=65.3 \pm 2.2$ km/s/Mpc at 68\% CL, for the different baselines adopted here. These results are at more than 4$\sigma$ tension with the local measurement. Also, we do not find any deviations from $\xi_{\nu} \neq 0$ using CMB data. The upper limits at 68\% CL on $\xi_{\nu}$ are reported in Table \ref{tab:CMB_FS}.

We now discuss the results from the CMB + FS dataset combination, that is,  by considering FS power spectrum measurements for the BOSS and eBOSS data with the CMB data. First, taking the baseline-model with $N_{\rm eff}=3.046$ and $\sum m_{\nu} =0.06 $ eV, i.e., $\Lambda$CDM + $c^2_{\rm  eff}$ + $c^2_{\rm  vis}$ + $\xi_{\nu}$ model, we note a significant improvement in the constraints by the introduction of the FS dataset on the full base parameters, viz., CMB + eBOSS and CMB + BOSS data improve the constraints on the model baseline, with the exception for the bounds on $\xi_{\nu}$. Thus, there is an increasing order of robustness with CMB, CMB + eBOSS and CMB + BOSS data, respectively, see Table \ref{tab:CMB_FS}, Fig. \ref{fig:cmb_FS}. Therefore, clearly the addition of FS measurements improves the constraints.  It is important to note that BOSS sample covers positions and redshifts of four independent data-sets, while eBOSS sample does only two, as previously defined in Section \ref{data}. Thus, naturally, the analysis with BOSS generates more robust results than eBOSS, given the effective volumes of each sample in consideration.

Now, considering our second baseline-model, we can still notice 
some other perspectives. But, first we should mention that when $N_{\rm eff}$ and $\sum m_{\nu}$ become free parameters, naturally the error bars on the other parameters increase, and also the degree of correlations increases in comparison with the previous common baseline. In this specific model-baseline, when compared with the CMB only analysis, it becomes more evident that FS measurements can improve significantly the constraints on the full parametric space. We notice significant improvement on all parameters in the analysis with the CMB + BOSS data. The constraints on the parameters $c^2_{\rm eff}$, $c^2_{\rm vis}$ and $N_{\rm eff}$, which quantify the neutrinos properties, are remarkably improved, see Figure \ref{fig:cmb_FS_neff}. In particular, we find $\sum m_{\nu} < 0.413$ eV at 95\% CL from CMB only, and $\sum m_{\nu} < 0.116$ eV at 95\% CL from CMB + BOSS. It represents $\sim$28\% improvement in comparion with CMB only case. Within this generalized framework, we find $H_0=67.5 \pm 1.2$ km/s/Mpc and $S_8=0.821 \pm 0.013$, both at 68\% CL, from CMB + BOSS, which is a very robust constraint within this extended $\Lambda$CDM parametric space. Again, we do not find any evidence beyond the default values for all neutrino properties. The constraints from CMB + eBOSS, in terms of accuracy, are intermediate between CMB only and CMB + BOSS results. See Figure \ref{fig:cmb_FS_neff} and Table \ref{tab:CMB_FS}.

As previously mentioned, the addition of FS power spectrum measurements improve the observational constraints considerably on $\xi_{\nu}$, $H_0$ and $S_8$, $N_{\rm eff}$ and $\sum m_{\nu}$,  $c^2_{\rm  eff}$, $c^2_{\rm  vis}$. The effect of the parameters $c^2_{\rm eff}$, $c^2_{\rm vis}$ on the power spectrum was discussed in detail in \cite{Audren_2015}. This non-standard model has effects below 1\% on large scales (for $k <$ 0.01 Mpc$^{-1}$), but on smaller scales the effects can become significant for $c^2_{\rm eff}$. At scales between 0.01 and 0.2 Mpc$^{-1}$ increasing (decreasing) any of the two sound speed parameters causes a decrease (increase) on the amplitude of the power spectrum. This amplitude modulation is still below 1\%, even for significant changes in $c^2_{\rm vis}$, but $c^2_{\rm eff}$ can introduce modulations on the shape of the power spectrum up to quasi-non-linear and mainly on non-linear scale. Therefore, within the limits of analysis in this work from the BOSS and eBOSS data, we are not able to robustly constrain these specific parameters using the FS galaxy power spectrum data only, especially $c^2_{\rm vis}$. On the other hand, as demonstrated and explored in several previous works, the FS data alone are not very sensitive to $\omega_b$, $n_s$ and $\sum m_{\nu}$. In particular, the BOSS dataset (same considerations for eBOSS data) itself can only rule out very large neutrino masses $\sim$ 1 eV (see \cite{Ivanov:2019hqk,Colas:2019ret} and references therein), which produce significant scale-dependent modifications to the matter power spectrum. Smaller neutrino masses  can not be constrained with the BOSS data (FS data in general). The FS data can only improve the current neutrino mass bounds by breaking degeneracies internal to the CMB data. Same conclusion holds for $c^2_{\rm vis}$ and $c^2_{\rm eff}$.

Taking these considerations, we consider the BOSS + eBOSS + BBN joint analysis for the $\Lambda$CDM + $c^2_{\rm eff}$ +  $c^2_{\rm vis}$ + $\xi_{\nu}$ model. Here, BBN means a Gaussian prior on the physical baryon density parameter $\omega_b$ arising from Big Bang Nucleosynthesis (BBN) constraints on the abundance of light elements: $100\omega_b =  2.233 \pm 0.036$~\cite{Mossa:2020gjc}. Note that, in principle, the FS data can constrain $\omega_b$ without any external input. However, this constraint is expected to be much weaker than the measurements from BBN or Planck. Thus, keeping in mind the eventual combination of the FS and Planck data as presented previously, it is reasonable to impose this prior.  As expected, the parameters $c^2_{\rm eff}$, $c^2_{\rm vis}$, $\xi_{\nu}$ are weakly constrained. We find $c^2_{\rm eff} < 0.36$ at 68\% CL. The parameters $c^2_{\rm vis}$ and $\xi_{\nu}$ are practically unconstrained within some reasonable physical prior  range. In particular, it would be necessary to go into the deeply nonlinear regime to constraint $c^2_{\rm vis}$ parameter with only FS information. For BOSS + eBOSS + BBN data, we note that $\Omega_m$ is well constrained, viz., $\Omega_m=0.308^{+0.026}_{-0.029}$ at 68\% CL (and $S_8=0.751^{+0.077}_{-0.089}$ at 68\% CL). Therefore the amplitude of perturbations are lower than predicted by the minimum standard model from the CMB analysis, but statistically compatible with each other. For the Hubble parameter, we find the constraint $H_0=72.2^{+3.2}_{-4.2}$ km/s/Mpc at 68\% CL.  Therefore, without CMB data, information from LSS alone (the mildly non-linear full-shape galaxy power spectrum) can not provide good accuracy on these neutrinos properties, as well as on the full model baseline, in general.
We mention that FS clustering analyses in the minimal $\Lambda$CDM model with BOSS + eBOSS were recently carried out in \cite{Semenaite, Neveux}. These works also take into account the quasar sample, which is not included in our analysis.

\begin{table*}[!t]
\caption{Marginalized constraints on the bias parameters at 68\% CL extracted from BOSS and eBOSS galaxies samples used in this work for the model-baseline, $\Lambda$CDM + $c^2_{\rm  eff}$ + $c^2_{\rm  vis}$ + $\xi_{\nu}$.  from the NGC and SGC respectively.}
\centering
\scalebox{0.75}{
\begin{tabular}{|c|cc|cc|cc|cc}       
\hline
Parameter & BOSS $z_{\rm eff}=0.38$ && BOSS $z_{\rm eff}=0.61$ && eBOSS $z_{\rm eff}=0.698$ &  \\ \hline
          & NGC & SGC & NGC & SGC &  NGC & SGC  \\ \hline

$b_1$ & $1.876\pm 0.042            $ & $1.871^{+0.056}_{-0.050}   $  & $1.983\pm 0.045            $ & $2.025\pm 0.050            $ & $2.244\pm 0.059            $ & $2.310^{+0.061}_{-0.053}   $ \\
$b_2$ &  $-0.63^{+0.66}_{-0.78}     $ &  $-0.58^{+0.67}_{-0.94}     $ &$-1.16^{+0.71}_{-0.90}     $ & $-0.29^{+0.81}_{-0.97}     $ & $-2.01^{+0.68}_{-0.88}     $ & $-1.04^{+0.77}_{-0.97}     $   \\
$b_{G 2}$ & $-0.142^{+0.096}_{-0.12}   $ & $0.21^{+0.15}_{-0.18}      $  & $-0.09^{+0.14}_{-0.17}     $ & $0.08\pm 0.20              $ & $-0.31^{+0.11}_{-0.18}     $ & $-0.21\pm 0.14             $  \\

\hline                                                
\end{tabular}
}
\label{tab:bias_1}
\end{table*}

\begin{table*}[!t]
\centering
\caption{Same as Table \ref{tab:bias_1}, but for the model: $\Lambda$CDM + $N_{\rm eff}$ + $\sum m_{\nu}$ + $c^2_{\rm  eff}$ + $c^2_{\rm  vis}$ + $\xi_{\nu}$  }
\scalebox{0.75}{
\begin{tabular}{|c|cc|cc|cc|cc}       
\hline
Parameter &  BOSS $z_{\rm eff}=0.38$ && BOSS $z_{\rm eff}=0.61$ && eBOSS $z_{\rm eff}=0.698$ &  \\ \hline
          & NGC & SGC & NGC & SGC &  NGC & SGC  \\ \hline
$b_1$ & $1.876\pm 0.046            $ & $1.870\pm 0.056            $  & $1.979\pm 0.047            $ & $2.028\pm 0.055            $ &  $2.266\pm 0.073            $ &  $2.337\pm 0.072            $ \\
$b_2$ & $-0.65^{+0.67}_{-0.83}     $ & $-0.56^{+0.72}_{-0.93}     $ & $-1.29^{+0.67}_{-0.89}     $ & $-0.34^{+0.85}_{-0.99}     $ & $-1.97^{+0.68}_{-0.99}     $ & $-1.07^{+0.79}_{-0.95}     $  \\
$b_{G 2}$ &  $-0.142^{+0.095}_{-0.12}   $ & $0.20^{+0.15}_{-0.18}      $  & $-0.09^{+0.14}_{-0.18}     $ &$0.08\pm 0.21              $ & $-0.34^{+0.11}_{-0.17}     $ & $-0.23\pm 0.15             $  \\

\hline                                                
\end{tabular}
}
\label{tab:bias_2}
\end{table*}

\begin{figure*}[hbt!]
\begin{center}
\includegraphics[width=15cm]{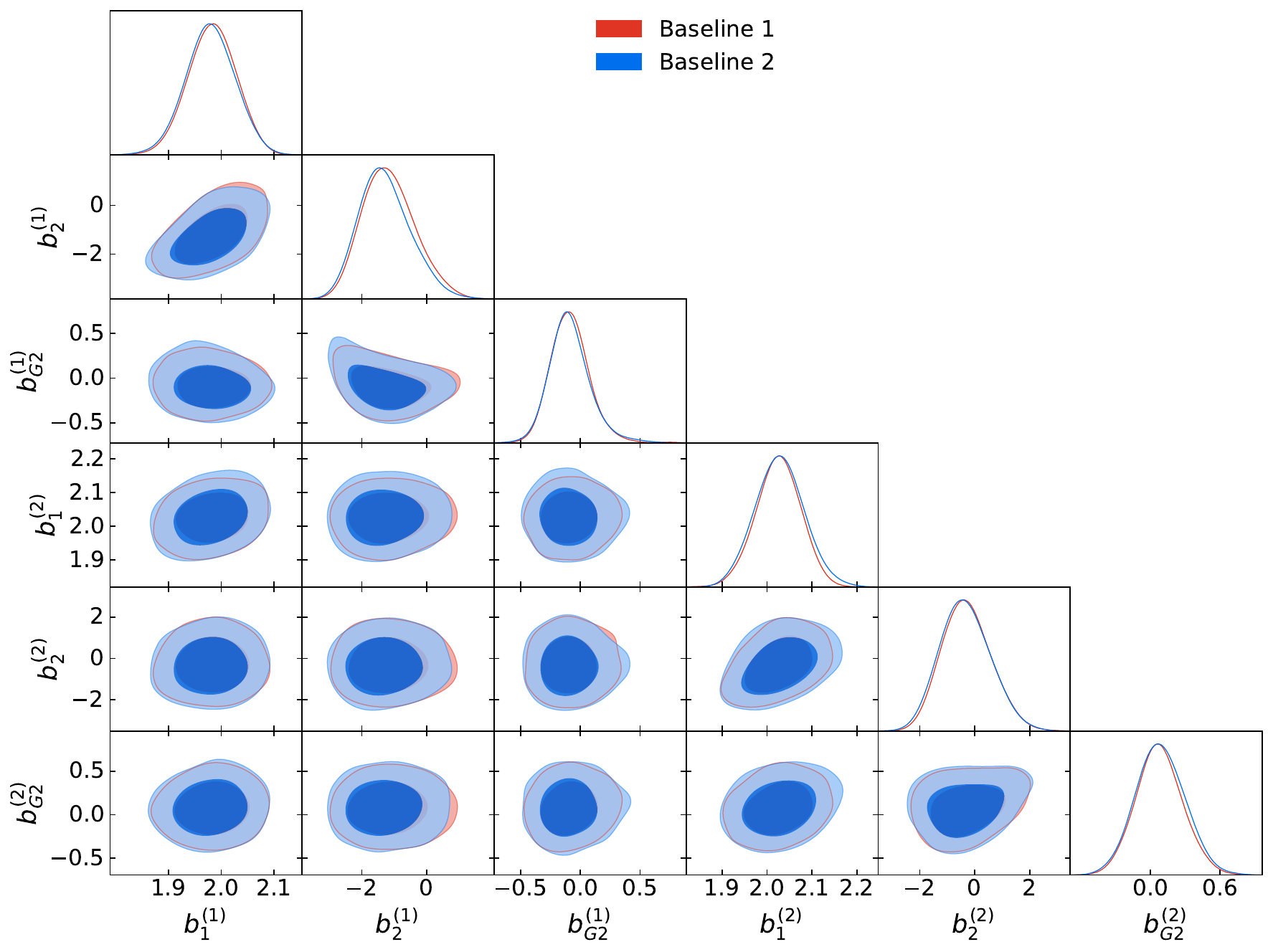}
\caption{One-dimensional and two-dimensional marginalized confidence regions (68\% and 95\% CL) for the bias parameters $b_1$, $b_2$ and $b_{G 2}$ for the BOSS data sample at effetive $z = 0.61$. The over-index $i=$ 1, 2 means the constraints from the NGC and SGC respectively. The red region (Baseline 1) represents the model with $N_{\rm eff}=3.046$ and $\sum m_{\nu}=0.06$ eV while the blue region (Baseline 2) with $N_{\rm eff}$ and $\sum m_{\nu}$ as free parameters, as discussed in the main text. We clearly see a perfect agreement of bias parameters between the different models.}
\label{fig:bias}
\end{center}
\end{figure*}

\begin{figure*}[hbt!]
\begin{center}
\includegraphics[width=15cm]{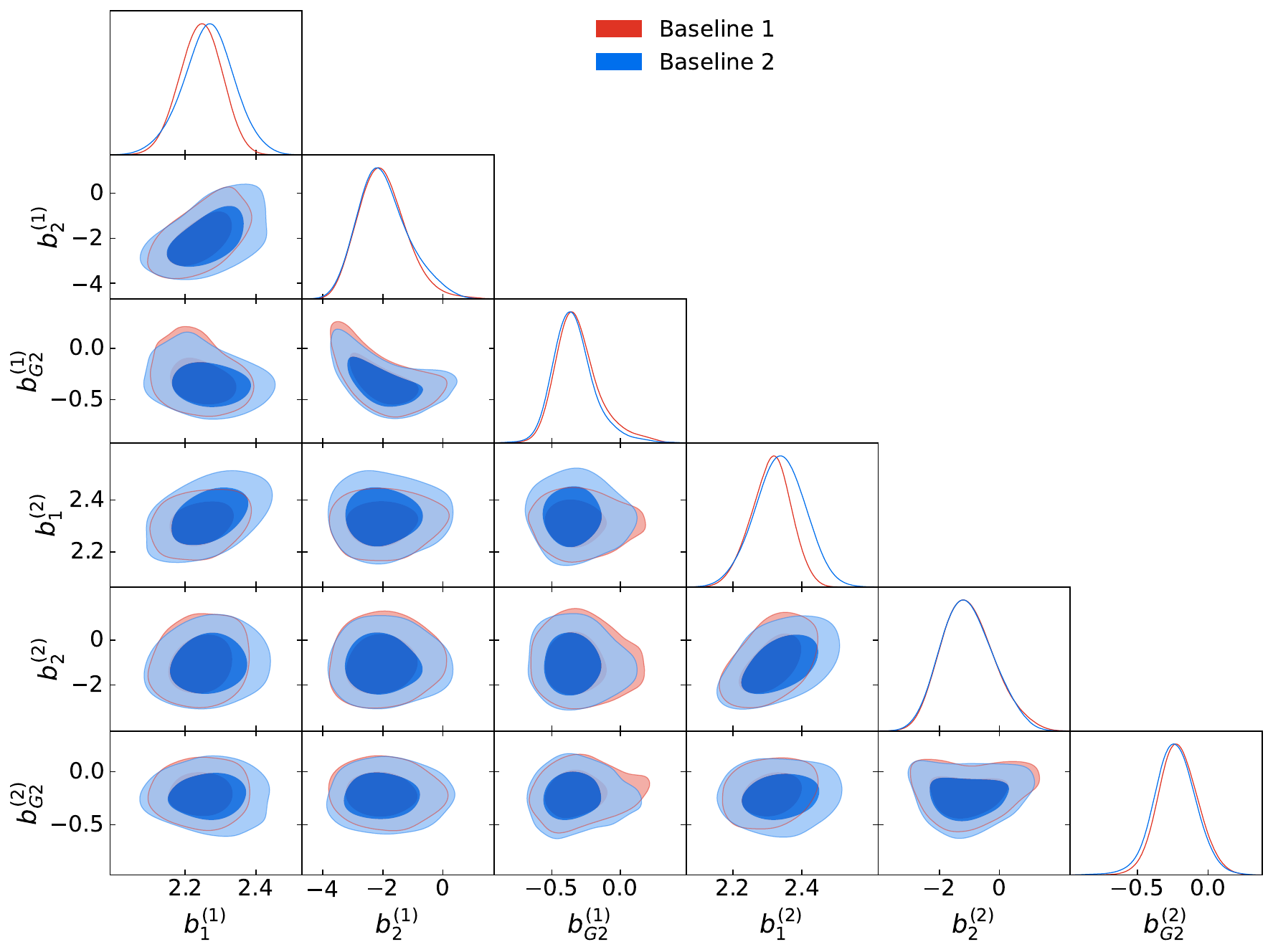} 
\caption{Same as Figure \ref{fig:bias}, but for eBOSS galaxies data sample.}
\label{fig:bias_eboss}
\end{center}
\end{figure*}

Finally, it is important to explore the impact of the models on the bias parameters. We check whether the model-baseline i) $\Lambda$CDM + $c^2_{\rm  eff}$ + $c^2_{\rm  vis}$ + $\xi_{\nu}$ and ii) $\Lambda$CDM + $N_{\rm eff}$ + $\sum m_{\nu}$ + $c^2_{\rm  eff}$ + $c^2_{\rm  vis}$ + $\xi_{\nu}$, may affect the bias parameters in the final fit of the data.  Table \ref{tab:bias_1} and \ref{tab:bias_2} summarize constraints on the bias parameters at 68\% CL extracted from BOSS and eBOSS galaxies samples used in this work. Figure \ref{fig:bias} shows the one-dimensional and two-dimensional marginalized confidence regions (68\% and 95\% CL) of the bias parameters $b_1$, $b_2$ and $b_{G 2}$ from the BOSS data sample at effective redshift $z = 0.61$. The over-index $i=$ 1, 2 means the constraint from the NGC and SGC respectively.  We find no deviations on the bias parameters in  the extended $\Lambda$CDM models investigated in this work. Therefore, the bias parameters are self-consistent across different models. We come to the same conclusion for the sample at $z_{\rm eff}=0.38$. In Figure \ref{fig:bias_eboss}, we quantify the same perspective and conclusion for the eBOSS data at $z_{\rm eff}=0.698$.

\section{Final Remarks}
\label{conclusions}

We have updated and improved the constraints on the neutrino properties within  two extended $\Lambda$CDM scenarios using Planck-CMB 2018 data and its combination with the pre-reconstructed FS galaxy power spectrum measurements from the BOSS DR12 sample and eBOSS LRG DR16 sample. By exploiting the information content of the FS power spectrum of LSS tracers, we have shown that the combination CMB + FS  significantly improves the constraint on non-standard neutrinos properties. More specifically, this joint analysis  improves the previously known results up to an order of magnitude. Any deviation in the baseline $\{N_{\rm eff}, c^2_{\rm eff}, c^2_{\rm vis}, \xi_{\nu}  \}$ would mean evidence for a new physics, but our analysis does not report such an evidence, and our results are consistent with the values predicted by standard cosmology theory. Our most robust observational constraints are given by the joint analysis CMB + BOSS summarized in Table \ref{tab:CMB_FS}, which provide the strongest limits ever reported for the extended scenario. 

In this work,  we have utilized the FS measurements from the BOSS DR12 sample and  eBOSS LRG DR16 sample. Further, it would be interesting to forecast the model with the future LSS surveys, such as Euclid \cite{Amendola_2018} and
DESI \cite{DESI}. Also, it would be interesting to confront the model with N-body simulations to constrain $c^2_{\rm eff}$ and $c^2_{\rm vis}$ with greater precision using only the LSS information. Finally, it could be worthwhile to test other well-motivated extended-$\Lambda$CDM models in light of the FS measurements. Most of these ideas are the subject of work in progress, which we hope to report soon.

\acknowledgments
The authors thank the referee for valuable suggestions and comments, also thank Sunny Vagnozzi for useful discussions. S.K. gratefully acknowledges support from the Science and Engineering Research Board (SERB), Govt. of India (File No.~CRG/2021/004658). R.C.N. acknowledges financial support from the Funda\c{c}\~{a}o de Amparo \`{a} Pesquisa do Estado de S\~{a}o Paulo (FAPESP, S\~{a}o Paulo Research Foundation) under the project No.~2018/18036-5. P.Y. is supported by Junior Research Fellowship (CSIR/UGC Ref. No. 191620128350) from University Grant Commission, Govt. of India.


\end{document}